\documentclass[9pt]{article}
\usepackage{spconf,amsmath,graphicx}
\usepackage[numbers]{natbib}
\usepackage{multirow}
\usepackage{textcomp}
\usepackage{graphicx,scalerel}
\usepackage{moredefs} % for use of \defcommand. NOTE: can replace \defcommand by \(re)newcommand, but \defcommand will auto reassign an existing command

\newcommand\wh[1]{\hstretch{2}{\hat{\hstretch{.5}{#1}}}}

\usepackage{color}
\definecolor{darkspringgreen}{rgb}{0.09, 0.45, 0.27}
\definecolor{bronze}{rgb}{0.8, 0.5, 0.2}

\title{Deliberation Model Based Two-Pass End-to-End Speech Recognition}
\name{Ke Hu, Tara N. Sainath, Ruoming Pang, Rohit Prabhavalkar}
\address{Google, Inc., USA \\
\fontsize{9}{9}\selectfont\ttfamily\upshape
\{huk,tsainath,rpang,prabhavalkar\}@google.com}
\begin{document}
\ninept
\maketitle
\begin{abstract}
End-to-end (E2E) models have made rapid progress in automatic speech recognition (ASR) and perform competitively relative to conventional models. To further improve the quality, a two-pass model has been proposed to rescore streamed hypotheses using the non-streaming Listen, Attend and Spell (LAS) model while maintaining a reasonable latency. The model attends to acoustics to rescore hypotheses, as opposed to a class of neural correction models that use only first-pass text hypotheses. In this work, we propose to attend to both acoustics and first-pass hypotheses using a deliberation network. A bidirectional encoder is used to extract context information from first-pass hypotheses. The proposed deliberation model achieves 12\% relative WER reduction compared to LAS rescoring in Google Voice Search (VS) tasks, and 23\% reduction on a proper noun test set. Compared to a large conventional model, our best model performs 21\% relatively better for VS. In terms of computational complexity, the deliberation decoder has a larger size than the LAS decoder, and hence requires more computations in second-pass decoding.
\end{abstract}
%
%\begin{keywords}
%Deliberation network, end-to-end speech recognition, attention-based models
%\end{keywords}
%
\section{Introduction}
\label{sec:intro}

E2E ASR has gained a lot of popularity due to its simplicity in training and decoding. An all-neural E2E model eliminates the need to individually train components of a conventional model (i.e., acoustic, pronunciation, and language models), and directly outputs subword (or word) symbols \cite{graves2012sequence,rao2017exploring,chan2016listen,bahdanau2016end,kim2017joint}.  In large scale training, E2E models perform competitively compared to more sophisticated conventional systems on Google traffic \cite{he19streaming, chiu18}. Given its all-neural nature, an E2E model can be reasonably downsized to fit on mobile devices \cite{he19streaming}.

Despite the rapid progress made by E2E models, they still face challenges compared to state-of-the-art conventional models \cite{pundak2016lower,luscher2019rwth}. To bridge the quality gap between a streaming recurrent neural network transducer (RNN-T) \cite{he19streaming} and a large conventional model \cite{pundak2016lower}, a two-pass framework has been proposed in \cite{sainath2019twopass}, which uses a non-streaming LAS decoder to \emph{rescore} the RNN-T hypotheses. The rescorer attends to audio encoding from the encoder, and computes sequence-level log-likelihoods of first-pass hypotheses. The two-pass model achieves 17\%-22\% relative WER reduction (WERR) compared to RNN-T \cite{he19streaming} and has a similar WER to a large conventional model \cite{pundak2016lower}. 

A class of neural correction models post-process hypotheses using only the text information, and can be considered as second-pass models  \cite{zhang2019neural,zhang2019investigation,guo2019spelling}. The models typically use beam search to generate new hypotheses, compared to rescoring where one leverages external language models trained with large text corpora \cite{kannan2018analysis}. For example, a neural correction model in \cite{zhang2019neural} takes first-pass text hypotheses and generates new sequences to improve numeric utterance recognition \cite{peyser2019improving}. A transformer-based spelling correction model is proposed in \cite{zhang2019investigation} to correct the outputs of a connectionist temporal classification model in Mandarin ASR. In addition, \cite{guo2019spelling} leverages text-to-speech (TTS) audio to train an attention-based neural spelling corrector to improve LAS decoding. These neural correction models typically use only text as inputs, while the aforementioned two-pass model attends to acoustics alone for second-pass processing. 

In this work, we propose to combine acoustics and first-pass text hypotheses for second-pass decoding based on the deliberation network \cite{xia2017deliberation}. The deliberation model has been used in state-of-the-art machine translation \cite{hassan2018achieving}, or generating intermediate representation in speech-to-text translation \cite{sung2019towards}. Our deliberation model has a similar structure as \cite{xia2017deliberation}: An RNN-T model generates the first-pass hypotheses, and deliberation attends to both acoustics and first-pass hypotheses for a second-pass \emph{decoding}. We encode first-pass hypotheses bidirectionally to leverage context information for decoding. Note that the first-pass hypotheses are sequences of wordpieces \cite{schuster2012japanese} and are usually short in VS, and thus the encoding should have limited impact on latency.

Our experiments are conducted using the same training data as in \cite{narayanan2018toward,narayanan2019recognizing}, which is from multiple domains such as Voice Search, YouTube, Farfield and Telephony. We first analyze the behavior of the deliberation model, including performance when attending to multiple RNN-T hypotheses, contribution of different attention, and rescoring vs. beam search. We apply additional encoder (AE) layers and minimum WER (MWER) training \cite{prabhavalkar2018minimum} to further improve quality. The results show that our MWER trained 8-hypothesis deliberation model performs 11\% relatively better than LAS rescoring \cite{sainath2019twopass} in VS WER, and up to 15\% for proper noun recognition. Joint training further improves VS slightly (2\%) but significantly for a proper noun test set: 9\%. As a result, our best deliberation model achieves a WER of 5.0\% on VS, which is 21\% relatively better than the large conventional model \cite{pundak2016lower} (6.3\% VS WER). Lastly, we analyze the computational complexity of the deliberation model, and show some decoding examples to understand its strength.

\section{Deliberation Based Two-Pass E2E ASR}
\label{sec:model}

\subsection{Model Architecture}
\label{sec:architecture}

\begin{figure}[h!]
  \centering
   \includegraphics[scale=0.41]{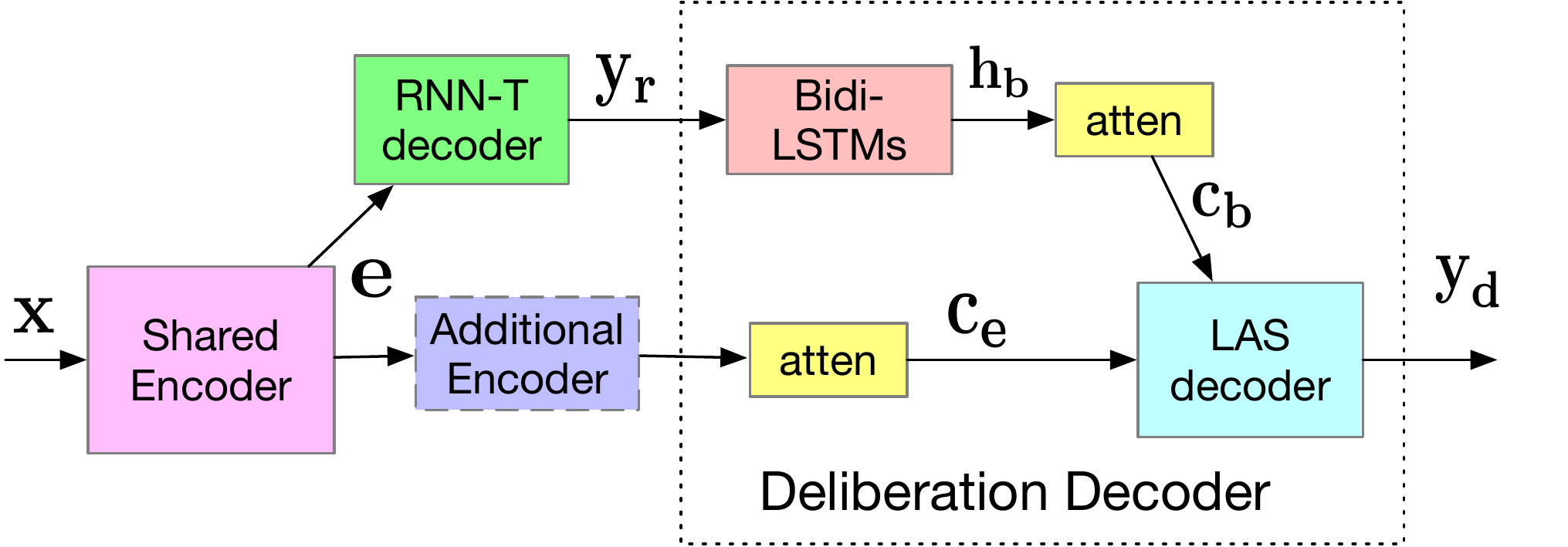}
   \caption{Diagram of the deliberation model with an optional additional encoder (dashed box).}
   \label{fig:delib}
\end{figure}

As shown in Fig. \ref{fig:delib}, our deliberation network consists of three major components: A shared encoder, an RNN-T decoder \cite{graves2012sequence}, and a deliberation decoder, similar to \cite{sainath2019twopass, xia2017deliberation}. The shared encoder takes log-mel filterbank energies, $\mathbf{x} = ( \mathbf{x}_1,...,\mathbf{x}_T )$, where $T$ denotes the number of frames, and generates an encoding $\mathbf{e}$. The encoder output $\mathbf{e}$ is then fed to an RNN-T decoder to produce first-pass decoding results $\mathbf{y}_r$ in a streaming fashion. Then the deliberation decoder attends to both $\mathbf{e}$ and $\mathbf{y}_r$ to predict a new sequence $\mathbf{y}_d$. We use a bidirectional encoder to further encode $\mathbf{y_r}$ for useful context information, and the output is denoted as $\mathbf{h}_b$. Note that we could use multiple hypotheses $\{\mathbf{y}_r^i\}$, where $i=1,...,H$ and $H$ is the number of hypotheses, and in this scenario we encode each hypothesis $\mathbf{y}_r^i$ separately using the same bidirectional encoder, and then concatenate their outputs in time to form $\mathbf{h}_b$. We keep the audio encoder unidirectional due to latency considerations. Then, two attention layers are followed to attend to acoustic encoding and first-pass hypothesis encoding separately. The two context vectors, $\mathbf{c}_b$ and $\mathbf{c}_e$, are concatenated as inputs to a LAS decoder.

There are two major differences between our model and the LAS rescoring \cite{sainath2019twopass}. First, the deliberation model attends to both $\mathbf{e}$ and $\mathbf{y}_r$, while \cite{sainath2019twopass} only attends to the acoustic embedding, $\mathbf{e}$.  Second, our deliberation model encodes $\mathbf{y}_r$ bidirectionally, while \cite{sainath2019twopass} only relies on unidirectional encoding $\mathbf{e}$ for decoding.

\subsubsection{Additional Encoder Layers}
\cite{sainath2019twopass} shows that the incompatibility between an RNN-T encoder and a LAS decoder leads to a gap between the rescoring model and LAS-only model. To help adaptation, we introduce a 2-layer LSTM as an additional encoder (dashed box in Fig. \ref{fig:delib} to indicate optional) to further encode $\mathbf{e}$.
We show in Sect. \ref{sec:results} that additional encoder layers improve both deliberation and LAS rescoring models.

\subsection{Training}
\label{sec:training}

A deliberation model is typically trained from scratch by jointly optimizing all components \cite{xia2017deliberation}. However, we find training a two-pass model from scratch tends to be unstable in practice \cite{sainath2019twopass}, and thus use a two-step training process: Train the RNN-T as in \cite{he19streaming}, and then fix the RNN-T parameters and only train the deliberation decoder and additional encoder layers as in \cite{chiu18,sainath2019twopass}.

\subsubsection{MWER Loss}
\label{sec:mwer_loss}

We apply the MWER loss \cite{prabhavalkar2018minimum} in training which optimizes the expected word error rate by using n-best hypotheses:
\begin{equation}
L_{\text{MWER}}(\mathbf{x}, \mathbf{y}^*) = \sum_{i=1}^{B} \hat{P}(\mathbf{y}_d^i|\mathbf{x})[W(\mathbf{y}_d^i, \mathbf{y}^*) - \wh{W}]
\label{eq:mwer}
\end{equation}
\noindent
where $\mathbf{y}_d^i$ is the $i$th hypothesis from the deliberation decoder, and $W(\mathbf{y}_d^i, \mathbf{y}^*)$ the number of word errors for $\mathbf{y}_d^i$ w.r.t the ground truth target $\mathbf{y}^*$. $\hat{P}(\mathbf{y}_d^i|\mathbf{x})$ is the probability of the $i$th hypothesis normalized over all other hypotheses to sum to 1. $B$ is the beam size. In practice, we combine the MWER loss with cross-entropy (CE) loss to stabilize training: $L'_{\text{MWER}}(\mathbf{x}, \mathbf{y}^*)=L_{\text{MWER}}(\mathbf{x}, \mathbf{y}^*) + \alpha L_{\text{CE}}(\mathbf{x}, \mathbf{y}^*)$, where $\alpha=0.01$ as in \cite{prabhavalkar2018minimum}.

\subsubsection{Joint Training}
\label{sec:joint_training}

Training the deliberation decoder while fixing RNN-T parameters is not optimal since the model components are not jointly updated. We propose to use a combined loss to train all modules jointly:

\begin{equation}
L_{\text{joint}}(\theta_e, \theta_1, \theta_2) = L_{\text{RNNT}}(\theta_e, \theta_1) + \lambda L_{\text{CE}}(\theta_e, \theta_2) 
\label{eq:joint_loss}
\end{equation}
\noindent
where $L_{\text{RNNT}}(\cdot)$ is the RNN-T loss, and $L_{\text{CE}}(\cdot)$ the CE loss for the deliberation decoder. $\theta_e$, $\theta_1$, and $\theta_2$ denote the parameters of shared encoder, RNN-T decoder, and deliberation decoder, respectively. Note that a jointly trained model can be further trained with MWER loss. The joint training is similar to ``deep finetuning" in \cite{sainath2019twopass} but without a pre-trained decoder.

\subsection{Decoding}
Our decoding consists of two passes: 1) Decode using the RNN-T model to obtain the first-pass sequence $\mathbf{y}_r$, and 2) Attend to both $\mathbf{y}_r$ and $\mathbf{e}$, and perform the second beam search to generate $\mathbf{y}_d$. We are also curious how rescoring performs given bidirectional encoding from $\mathbf{y}_r$. In rescoring, we run the deliberation decoder on $\mathbf{y}_r$ in a teacher-forcing mode \cite{sainath2019twopass}. Note the difference from \cite{sainath2019twopass}
when rescoring a hypothesis is that the deliberation network sees all candidate hypotheses. We compare rescoring and beam search in Sect. \ref{sec:results}.

\section{Experimental Setup}

\subsection{Datasets}
For training, we use the same multidomain datasets as in \cite{narayanan2018toward,narayanan2019recognizing} which include anonymized and hand-transcribed English utterances from general Google traffic, far-field environments, telephony conversations, and YouTube. We augment the clean training utterances by artificially corrupting them by using a room simulator, varying degrees of noise, and reverberation such that the signal-to-noise ratio (SNR) is between 0dB and 30dB \cite{kim2017generation}. We also use mixed-bandwidth utterances at 8kHz or 16 kHz for training \cite{yu2013feature}.

Our main test set includes \texttildelow14K anonymized hand-transcribed VS utterances sampled from Google traffic. To evaluate the performance of proper noun recognition, we report performance on a side-by-side (SxS) test set, and 4 voice command test sets \cite{he19streaming}. The SxS set contains utterances where the LAS rescoring model \cite{sainath2019twopass} performs inferior to a state-of-the-art conventional model \cite{pundak2016lower}, and one reason is due to proper nouns. The voice command test sets include 3 TTS test sets created using parallel-wavenet \cite{oord17}: Songs, Contacts-TTS, and Apps, where the commands include song, contact, and app names, respectively. The Contacts-Real set contains anonymized and hand-transcribed utterances from Google traffic to communicate with a contact, for example, ``Call Jon Snow".

\subsection{Architecture Details and Training}
Our first-pass RNN-T model has the same architecture as \cite{he19streaming}. The encoder of the RNN-T consists of an 8-layer Long Short-Term Memory (LSTM) \cite{hochreiter1997} and the prediction network contains 2 layers. Each LSTM layer has 2,048 hidden units followed by 640-dimensional projection. A time-reduction layer is added after the second layer to improve the inference speed without accuracy loss. Outputs of the encoder and prediction network are fed to a joint-network with 640 hidden units, which is followed by a softmax layer predicting 4,096 mixed-case wordpieces.

The deliberation decoder can attend to multiple hypotheses, and RNN-T hypotheses with different lengths are thus padded with end-of-sentence label $\langle\backslash\texttt{s}\rangle$ to a length of 120. Each subword unit in a hypothesis is then mapped to a vector by a 96-dimensional embedding layer, and then encoded by a 2-layer bidirectional LSTM encoder, where each layer has 2,048 hidden units followed by 320-dimensional projection.
Each of the two attention models is a multi-headed attention \cite{vaswani2017attention} with four attention heads. The two output context vectors are concatenated and fed to a 2-layer LAS decoder (2,048 hidden units followed by 640-dimensional projection per layer). The LAS decoder has a 4,096-dimensional softmax layer to predict the same mixed-case wordpieces \cite{schuster2012japanese} as the RNN-T.

For feature extraction, we use 128-dimensional log-Mel features from 32-ms windows at a rate of 10 ms. Each feature is stacked with three previous frames to form a 512-dimensional vector, and then downsampled to a 30-ms frame rate. Our models are trained in Tensorflow \cite{abadi2016tensorflow} using the Lingvo framework \cite{shen2019lingvo}  on 8$\times$8 Tensor Processing Units (TPU) slices with a global batch size of 4,096.

\subsection{Computational Complexity}
We estimate the computational complexity of the deliberation decoder using the number of floating-point operations (FLOPS) required:
\begin{equation}
\texttt{FLOPS} = M_{B} \cdot N \cdot H + M_{D} \cdot N \cdot B + \texttt{FLOPS}_{\texttt{atten}}
\label{eq:ops}
\end{equation}

where $M_B$ is the size of the bidirectional encoder, $N$ the number of decoded tokens, and $H$ the number of first-pass hypotheses. $M_D$ denotes the size of the LAS decoder, and $B$ the second beam search size. $\texttt{FLOPS}_{\texttt{atten}}$ is the FLOPS required for two attention layers, and we compute it as the sum of multiplying the sizes of source and query matrices with the number of time frames and $N$, respectively.
Our deliberation decoder contains roughly 66M parameters, where the size of the bidirectional encoder is $M_B=22$M, LAS decoder is $M_D=42$M, and attention layers have 2M parameters.

\section{Results}
\label{sec:results}
In this section we analyze the importance of certain components of the deliberation model by ablation studies, improve the model by MWER and AE layers, and select one of our best deliberation models for comparison.

\subsection{Number of RNN-T Hypotheses}

The deliberation decoder may attend to multiple first-pass hypotheses. We encode the hypotheses separately, and then concatenate them as the input to the attention layer. We use a beam size of $8$ for RNN-T decoding. Unless stated otherwise, the WER we report is for VS test set. The third row in Table \ref{tab:wer_num_hyp} shows that the WER improves slightly when increasing the number of RNN-T hypotheses from 1 to 8. However, after applying MWER training, the WER improves continuously: 5.4\% to 5.1\%. We suspect that MWER training specifically helps deliberation attend to relevant parts of first-pass hypotheses. Since 8-hypothesis model gives the best performance, we use that for subsequent experiments. MWER training is not used for simplicity.

\begin{table}[h]
\centering
\begin{tabular}{ |c|c|c|c|c| }
    \hline
    ID & E1 & E2 & E3 & E4 \\ \hline\hline
    Model & 1 hyp & 2 hyp & 4 hyp & 8 hyp \\ \hline
    Deliberation & 5.5 & 5.4 & 5.4 & 5.4 \\ \hline
    + MWER & 5.4 & 5.3 & 5.2 & 5.1 \\ \hline
\end{tabular}
\caption{WERs (\%) of deliberation with different number of RNN-T hypotheses.}
\label{tab:wer_num_hyp}
\end{table}

\subsection{Acoustics vs. Text}

We are curious about how different attention ($\mathbf{c}_b$ vs $\mathbf{c}_e$) contribute to deliberation, and thus train separate models where we attend to either acoustics (\texttt{E5}) or text (\texttt{E6}) alone in training and inference. Table \ref{tab:acoustics_text} shows that either \texttt{E5} or \texttt{E6} perform significantly better than the baseline RNN-T model (\texttt{B0}) with a 9\% WERR. By using both attentions (\texttt{E4}), the model gains another 11\% relative improvement. It seems surprising that \texttt{E6} performs equally to \texttt{E5}. We note this could be because \texttt{E6} has a bidirectional encoder while \texttt{E5} does not.

\begin{table}[h!]
\centering
\begin{tabular}{ |c||c|c| }
    \hline
    ID & Model & WER (\%) \\ \hline
    B0 & RNN-T & 6.7 \\ \hline
    E5 & Acoustics alone & 6.1 \\ \hline 
    E6 & Text alone & 6.1 \\ \hline
    E4 & Both & 5.4 \\ \hline
\end{tabular}
\caption{WERs (\%) of baseline RNN-T model and deliberation models with different attention setup.}
\label{tab:acoustics_text}
\end{table}

\subsection{Additional Encoder Layers}
\label{sec:ae}

To help the deliberation decoder better adapt to the shared encoder, we add AE layers for dedicated encoding for the deliberation decoder. The AE consists of a 2-layer LSTM with 2,048 hidden units followed by 640-dimensional projection per layer. Beam search is used for decoding. In Table \ref{tab:ae}, we show that with AE layers (\texttt{E7}) the model performs around 4\% better than without (\texttt{E4}). Similarly, we apply AE to the LAS beam search (\texttt{B1}$\rightarrow$\texttt{B2}), and obtain similar improvements.

\begin{table}[h]
\centering
\begin{tabular}{ |c|c|c|c| }
    \hline
    ID & Model & WER \\ \hline\hline
    E7 & E4 + AE & 5.2 \\ \hline
    B1 & LAS & 6.1 \\ \hline
    B2 & LAS + AE & 5.8 \\ \hline
\end{tabular}
\caption{WERs (\%) with or without AE layers.}
\label{tab:ae}
\end{table}

\begin{table*}
\centering
\begin{tabular}{ |c||l|l|c|c|c|c|c|c|c| }
    \hline
    \multirow{2}{*}{ID} & \multirow{2}{*}{Model} & \multirow{2}{*}{Decoding} & \multicolumn{6}{|c|}{WER (\%)} & \multirow{2}{*}{\shortstack{Estimated\\ GFLOPS}} \\ \cline{4-9}
     & & & VS & SxS & Songs & Contacts-Real & Contacts-TTS & Apps & \\ \hline\hline
    B0 & RNN-T & Beam search & 6.7 & 35.2 & 11.9 & 15.9 & 24.3 & 7.8 & 3.5 \\ \hline
    B4 & LAS \cite{sainath2019twopass} & Rescoring & 5.7 & 31.4 & 10.9 & 14.7 & 22.6 & 7.5 & 4.8 \\ \hline
    B5 & LAS \cite{sainath2019twopass} & Beam search & 
    \textbf{5.5} & \textbf{29.0} & \textbf{11.7} & \textbf{14.7} & \textbf{22.9} & \textbf{8.3} & \textbf{4.8} \\ \hline
    E9 & Deliberation & Beam search & 5.1 & 26.6 & 9.9 & 13.7 & 22.3 & 7.1 & 8.8 \\ \hline
    E10 & \hspace{0.5em}+ Joint training & Beam search & \textbf{5.0} & \textbf{24.3} & \textbf{9.6} & \textbf{13.4} & \textbf{22.0} & \textbf{6.4} & \textbf{8.8} \\ \hline 
\end{tabular}
\caption{Comparison of RNN-T, LAS two-pass and deliberation models in WERs (\%) and GFLOPS. LAS two-pass and deliberation models are augmented with AE layers. All models are trained with MWER loss except the RNN-T model.}
\label{tab:wer}
\end{table*}

\subsection{Rescoring}
We propose to use the deliberation decoder to rescore first-pass RNN-T results, and expect bidirectional encoding to help compared to LAS rescoring \cite{sainath2019twopass}. Table \ref{tab:rescoring} shows that the deliberation rescoring (\texttt{E8}) performs 5\% relatively better than LAS rescoring (\texttt{B3}). AE layers are added to both models.

\begin{table}[h]
\centering
\begin{tabular}{ |c||c|c| }
    \hline
    ID & Rescoring & WER (\%) \\ \hline
    E8 & Deliberation & 5.7 \\ \hline  
    B3 & LAS + AE & 6.0 \\ \hline
\end{tabular}
\caption{Comparison of rescoring models.}
\label{tab:rescoring}
\end{table}

\subsection{Comparisons}
\label{sec:comp}
From the above analysis, an MWER trained 8-hypothesis deliberation model with AE layers performs the best, and thus we use that for comparison below.

In Table \ref{tab:wer}, we compare deliberation models with an RNN-T \cite{he19streaming} and LAS rescoring model \cite{sainath2019twopass} in different recognition tasks including VS and proper noun recognition. We include two deliberation models: An MWER trained 8-hypothesis deliberation model with AE layers (\texttt{E9}), and a jointly trained version (\texttt{E10}). For LAS two-pass model, we add AE layers to the model in \cite{sainath2019twopass} and evaluate both rescoring (\texttt{B4}) and beam search (\texttt{B5}). We note that all models are MWER trained except the RNN-T model, which we find little improvement. First, we note that two-pass models perform substantially better than RNN-T (\texttt{B0}) in both VS task (15\%--25\% WERR) and rare word test sets (e.g. up to 30\% in \texttt{E10} for the SxS set). This confirms that second-pass decoding brings additional benefits. Second, the MWER trained 8-hypothesis deliberation model with AE layers (\texttt{E9}) performs significantly better than LAS rescoring (\texttt{B4}) or beam search (\texttt{B5}). When beam search is used for both of the deliberation and LAS models, the WERR is 7\% for VS, and 8\% for the SxS set. We observe significant improvements for voice command test sets too. Third, joint training (\texttt{E10}) brings an additional 2\% relative improvement for VS, 9\% for the SxS set, and uniform improvements for voice command test sets.

\begin{figure}[h]
  \centering
  \includegraphics[scale=0.28]{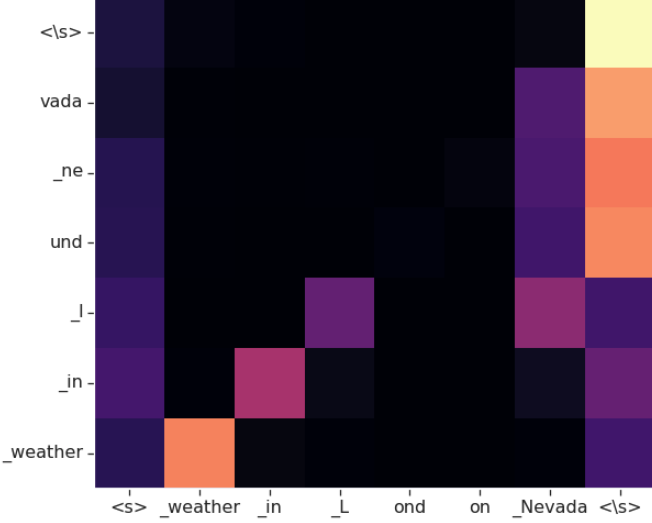}
  \caption{Example attention probabilities on a first-pass RNN-T hypothesis: ``\texttt{Weather in {\color{red}London} Nevada}", for generating the second-pass result ``\texttt{Weather in {\color{darkspringgreen}Lund} Nevada}". Brighter colors correspond to higher probabilities. A beginning wordpiece starts with a space marker (i.e., ``\textunderscore"). $\langle$\texttt{s}$\rangle$ denotes start of sentence, and $\langle$$\backslash$\texttt{s}$\rangle$ the end of sentence.}
  \label{fig:atten}
  \vspace{-1em}
\end{figure}

To understand where the improvement comes from, in Fig. \ref{fig:atten} we show an example of deliberation attention distribution on the RNN-T hypotheses (x-axis) at every step of the second-pass decoding (y-axis). We can see the attention selects mainly one wordpiece when the first-pass result is correct (e.g. ``\_\texttt{weather}", ``\_\texttt{in}", etc). However, when the first-pass output is wrong (e.g. ``\texttt{ond}" and ``\texttt{on}"), the attention looks ahead at ``\_\texttt{Nevada}" for context information for correction. We speculate that the attention functions similarly as a context-aware language model on the first-pass sequence.

In Table \ref{tab:wer}, we also report gigaFLOPS (GFLOPS) estimated using Eq. (\ref{eq:ops}) on the 90\%-tile VS set, where an utterance has roughly 109 audio frames and a decoded sequence of 14 tokens. Since the deliberation decoder has a larger size than LAS decoder (67MB vs. 33MB), it requires around 1.8 times GFLOPS as LAS rescoring. The increase mainly comes from the bidirectional encoder for 8 first-pass hypotheses. However, we note that the computation can be parallelized across hypotheses \cite{sainath2019twopass} and should have less impact on latency. Latency estimation is complicated, and we will quantify that in future works.

\subsection{Decoding Examples}

Lastly, we compare some decoding examples between deliberation and LAS rescoring in Table \ref{tab:decoding_example}. One type of wins for deliberation is URL, where the deliberation model corrects and concatenates string pieces to a single one since it sees the whole first-pass hypothesis. Second type is proper noun. Leveraging the context, deliberation realizes the previous word should be a proper noun (i.e. \texttt{Walmart}). Third, the deliberation decoder corrects semantic errors (\texttt{china} $\rightarrow$ \texttt{train}). On the other hand, we also see some losses of deliberation due to over-correction of proper nouns or spelling difference. The former is probably from knowledge in training, and the latter is benign and does not affect semantics.

\begin{table}[h]
\centering
\begin{tabular}{ |c|c| }
    \hline
     LAS rescoring & Deliberation \\ \hline
    % \multirow{3}{*}{Wins}
      {\color{red} Quality times.com} & {\color{darkspringgreen} quadcitytimes.com}  \\ \hline
      {\color{red} Where my} job application & {\color{darkspringgreen} Walmart} job application \\ \hline
    %   {\color{red} Runs a} casserole & {\color{darkspringgreen} runza} casserole \\ \hline
      {\color{red} china} near me & {\color{darkspringgreen} train} near me \\ \hline
      bio of {\color{darkspringgreen} Chesty} {\color{red} Fuller} & bio of {\color{red} Chester Fuller} \\ \hline
      2016 Kia {\color{darkspringgreen} Forte5} & 2016 Kia {\color{red} Forte 5} \\ \hline
\end{tabular}
\caption{Decoding examples of deliberation and LAS rescoring. Deliberation wins are in green and losses in red.}
\label{tab:decoding_example}
\end{table}

\section{Conclusion}
We presented a new two-pass E2E ASR based on the deliberation network, and our best model obtained significant improvements over LAS rescoring in both VS tasks and proper noun recognition: 12\% and 23\% WERR, respectively. The model also performs 21\% relatively better than a large conventional model for VS. Although the model requires more computation than LAS rescoring, batching across hypotheses can improve latency.

\renewcommand{\bibsection}{\section{REFERENCES}}
\bibliographystyle{abbrvunsrtnat}
{\footnotesize\bibliography{refs}}

\end{document}